%% file: main.tex
\documentclass[conference,compsoc]{IEEEtran}

\usepackage{tikz}
\usepackage{amsmath}

\usepackage{filecontents}

\usepackage{adjustbox}
\usepackage{listings}
\usepackage{booktabs}
\usepackage{multirow}
\usepackage{graphicx}
\usepackage{xcolor}
\usepackage{makecell}
\usepackage{tcolorbox}
\usepackage{multicol}
\usepackage{array}
\usepackage{tabularx}
\usepackage{tabulary}
\usepackage{svg}
\usepackage{comment}
\usepackage{url}

\usepackage{enumitem}
\setlist{leftmargin=5mm}  

\usepackage{titlesec}
\titlespacing\section{0pt}{6pt plus 3pt minus 1pt}{1pt plus 3pt minus 1pt}
\titlespacing\subsection{0pt}{6pt plus 3pt minus 1pt}{1pt plus 2pt minus 1pt}
\titlespacing\subsubsection{0pt}{6pt plus 3pt minus 1pt}{1pt plus 2pt minus 1pt}

\usepackage{pifont}%

\usepackage{hyperref}

\usepackage{multirow}
\usepackage{makecell}
\usepackage{dingbat}
\usepackage{booktabs}
\usepackage{geometry}
\usepackage{breakurl}

\definecolor{darkpastelgreen}{rgb}{0.01, 0.75, 0.24}

\definecolor{asparagus}{rgb}{0.53, 0.66, 0.42}
\definecolor{applegreen}{rgb}{0.55, 0.71, 0.0}
\definecolor{darkpastelgreen}{rgb}{0.01, 0.75, 0.24}
\definecolor{darkspringgreen}{rgb}{0.09, 0.45, 0.27}

\definecolor{darkpastelgreen}{rgb}{0.01, 0.75, 0.24}
\definecolor{cadmiumgreen}{rgb}{0.0, 0.42, 0.24}
\definecolor{codepurple}{rgb}{0.58,0,0.82}

\begin{document}

\date{}

\title{Open Source, Open Threats? Investigating Security Challenges in Open-Source Software}

\author{
    \IEEEauthorblockN{Seyed Ali Akhavani\textsuperscript{*}}
    \IEEEauthorblockA{\textit{Northeastern University} \\
    Boston, MA, USA \\
    sadatakhavani.s@northeastern.edu}
\and
    \IEEEauthorblockN{Behzad Ousat\textsuperscript{*}}
    \IEEEauthorblockA{\textit{Florida International University} \\
    Miami, FL, USA \\
    bousat@fiu.edu}
    
\and
    \IEEEauthorblockN{Amin Kharraz}
    \IEEEauthorblockA{\textit{Florida International University} \\
    Miami, FL, USA \\
    ak@cs.fiu.edu}
\and

\IEEEauthorblockN{\footnotesize \textsuperscript{*}Authors contributed equally to this work.}
}

\maketitle

\thispagestyle{plain}
\pagestyle{plain}

\input{arxiv/00-Abstract}
\begin{IEEEkeywords}
Supply Chain Security, Open-Source Security
\end{IEEEkeywords}

\input{arxiv/01-Introduction}

\input{arxiv/02-2-RelatedWork}

\input{arxiv/03-Methodology}

\input{arxiv/04-Data_Collection}

\input{arxiv/05-Evaluation}

\input{arxiv/06-Discussion}

\input{arxiv/90-Ethics}
\newpage
\bibliographystyle{ieeetr}
\bibliography{references}

\appendix

\input{arxiv/99-Appendix}

\end{document}

%% file: arxiv/00-Abstract.tex
\begin{abstract}
Open-source software (OSS) has become increasingly more popular across different domains. 
However, this rapid development and widespread adoption come with a security cost. The growing complexity and openness of OSS ecosystems have led to increased exposure to vulnerabilities and attack surfaces.
This paper investigates the trends and patterns of reported vulnerabilities within OSS platforms, focusing on the implications of these findings for security practices. To understand the dynamics of OSS vulnerabilities, we analyze a comprehensive dataset comprising 31,267 unique vulnerability reports from GitHub's advisory database and Snyk.io, belonging to 14,675 packages across 10 programming languages.

Our analysis reveals a significant surge in reported vulnerabilities, increasing at an annual rate of 98\%—far outpacing the 25\% average annual growth in the number of open-source software (OSS) packages.
Additionally, we observe an 85\% increase in the average lifespan of vulnerabilities across ecosystems during the studied period, indicating a potential decline in security.
We identify the most prevalent Common Weakness Enumerations (CWEs) across programming languages and find that, on average, just seven CWEs are responsible for over 50\% of all reported vulnerabilities. We further examine these commonly observed CWEs and highlight ecosystem-specific trends.
Notably, we find that vulnerabilities associated with intentionally malicious packages comprise 49\% of reports in the NPM ecosystem and 14\% in PyPI—an alarming indication of targeted attacks within package repositories. We conclude with an in-depth discussion of the characteristics and attack vectors associated with these malicious packages.

\end{abstract}

%% file: arxiv/01-Introduction.tex
\vspace{-0.5em}
\section{Introduction}
The open-source software (OSS) landscape has seen exponential growth over the past decade, becoming a backbone for a wide range of industries, including aerospace~\cite{dantas2024asapy}, energy systems\cite{ghiasi2023comprehensive, costa2022formation}, finance~\cite{wright2023open, yang2023fingpt}, healthcare~\cite{tyler2023improving}, and government projects. As the adoption of OSS starts to dominate across different sectors, its inherent benefits—such as transparency, collaboration, and rapid innovation—are accompanied by important challenges. The open and decentralized nature of OSS, while promoting collaboration and innovation, also creates opportunities for adversaries to exploit vulnerabilities~\cite{vupgrade_open_source_double_edged_sword}, compromise supply chains~\cite{cybersaint_open_source_impact,hackernews2023pypi}, and introduce malicious code into widely-used libraries~\cite{sysdig_hidden_economy_open_source,phylum_reverse_shell,bleepingcomputer2023malicious}.

Government entities, policymakers, and regulators recognize the need for stronger security measures in the open-source environment. The recent United States \textit{Securing Open Source Software Act} represents a significant step toward mandating security improvements and creating frameworks for the protection of open-source libraries and dependencies~\cite{whitehouse_oss_bill, whitehouse_oss_bill_summary}. Similarly, United States' Cybersecurity and Infrastructure Security Agency (CISA) has published the \textit{Open Source Software Security Roadmap}, outlining its strategy to ensure a secure open-source ecosystem~\cite{cisa_oss_roadmap}.
These initiatives reflect growing concerns over the reliance on OSS in critical infrastructure, highlighting challenges like dependency visibility, large-scale vulnerability management, and human errors in maintaining packages.

Our work is guided by three primary research questions. 1)~How have vulnerabilities evolved in different ecosystems?, 2)~How long do vulnerabilities persist and spread across packages?, 3)~What are the unique vulnerability patterns in different ecosystems? To answer these questions, we generated a dataset of vulnerability reports from 2017 to January 2025. The data consists of over \textbf{31,267 reported vulnerabilities} from two advisory databases--GitHub Advisory Database and Snyk.io-- that cover \textbf{14,675 open-source repositories} publicly available on GitHub.
We also have reports of \textbf{4,456 malicious packages} whose source code is no longer available since they have been permanently removed.
We examine vulnerabilities across \textbf{10 programming languages}--C, C++, PHP, Rust, JavaScript (Node.JS), Go, Java, .Net, Python, Ruby--, and \textbf{eight distinct package managers}--PyPi, NPM, Composer, Crates, Go, Maven, NuGet, and RubyGems-- which are foundational to software and web development~\cite{geeksforgeeks,lambdatest,browserstack,scaler}.
This longitudinal analysis allows us to track the growth and distribution of vulnerabilities across programming languages.

In the following, we highlight some of the major findings
of this paper.

\textbf{The rapid adoption of open-source platforms is being overshadowed by a faster rise in reported vulnerabilities.}
The analysis reveals a substantial increase in the number of packages across open-source platforms, with an average annual growth rate of 25.16\%. However, this growth is accompanied by an alarming rise in reported vulnerable packages, which have increased at a much faster rate—98\% annually. This suggests that as package managers rapidly expand their number of packages, the security risks associated with them are growing disproportionately. For example, in 2022, there were 2,490 vulnerabilities reported for Maven and 1,593 for Composer. The sharp increase in vulnerabilities, particularly in high-growth platforms, highlights a critical security gap, as the rate of vulnerabilities is outpacing the platform's expansion, making open-source software an increasingly attractive target for attackers.

\textbf{The expanding scale of open-source ecosystems does not necessarily translate into improved security.} We analyze vulnerabilities at the package level and find that the time a vulnerability remains in the ecosystem before being fixed has increased by 95\% from 2017 to 2024. 
Additionally, we examine vulnerability concentration patterns to assess whether security issues stem from a small number of packages or are dispersed across the ecosystem. Our analysis reveals distinct trends—some platforms exhibit a high concentration of vulnerabilities in key packages, while others show a broader, more distributed spread.
For instance, Composer (4.92 vulnerabilities per vulnerable package) and C/C++ (3.96) have vulnerabilities concentrated in a few critical packages. In contrast, ecosystems such as NPM (1.75 vulnerabilities per package) and Go (1.70) show a more even distribution of vulnerabilities.

\textbf{The evaluation of CWEs across ecosystems reveals distinct vulnerability patterns shaped by both common threats and language-specific characteristics.}
Our results reveal key trends in vulnerability distribution, highlighting the prevalence of certain vulnerabilities and the impact of language-specific characteristics. Our data demonstrates that, on average of the investigated platforms, only seven CWEs were responsible for over 50\% of the vulnerability reports.
Notably, CWE-79 (Cross-Site Scripting) and CWE-22 (Path Traversal) pose a major threat in ecosystems like Composer, NPM, PyPI, and Maven, reflecting widespread security challenges in web applications.
On the other hand, some CWEs are deeply tied to language characteristics, such as CWE-122 (Heap-based Buffer Overflow) in C/C++ or CWE-1321 (Prototype Pollution) in JavaScript.
We specifically focus on CWE-506 (Embedded Malicious Code) that has emerged as a significant issue in platforms like NPM and PyPI, where 48.58\% and 13.92\% of all vulnerabilities, represent intentional attacks underscoring the growing concern over supply chain attacks.

We hope this work to raise awareness about the increasing need for systematic approaches to enhance visibility in open-source software.
We also hope that the insights provided in this paper will inspire new directions in developing automated tools that can effectively model vulnerabilities and augment developers' experience to build more robust code and test cases. We summarize our contributions as follows:
\begin{itemize}[itemsep=0pt, topsep=3pt, partopsep=0pt, parsep=0pt, left=5pt]
    \item We present a longitudinal study that explores open-source software package across 10 programming languages and eight distinct package managers from 2017 to 2025, measuring the growth rate of packages and vulnerabilities.
 
    \item We investigate the lifespan of vulnerabilities over the last years across the ecosystems and discuss the uneven distribution of vulnerabilities across packages, identifying ecosystems with concentrated or widespread risks.
    
    \item We discuss the vulnerability types that are common across multiple platforms and others that are specifically reported for an ecosystem, requiring specific mitigation strategies for each case.
    
    \item We provide a detail analysis of intentionally malicious packages observed mostly in NPM and PyPI ecosystems, showcasing their characteristics and attack vectors.
    
\end{itemize}

\noindent \textbf{Availability.} 
We provide the research artifacts, including datasets and analysis scripts for further research~\footnote{\url{https://github.com/sa-akhavani/oss-security}}. This dataset serves as a valuable resource for future research in the field, as no publicly available dataset includes such valuable information in a unified format.

\input{diagrams/tab1}

%% file: diagrams/tab1.tex
\begin{table*}[h]
    \centering
    \renewcommand*{\arraystretch}{1.2}
    \caption{Overview of research paper details on package vulnerability and dependency network analysis.}
    \label{tab:related_work}
    \resizebox{1\textwidth}{!}{%
    
    \begin{tabular}{l|c|c|c|c|cccccccccc}
    \toprule
    \multirow{2}{*}{Category} & \multirow{2}{*}{Reference} & \multirow{2}{*}{\makecell{Total \\\ Vulnerabilities}} & \multirow{2}{*}{\makecell{Total \\ Packages}} & \multirow{2}{*}{\makecell{Evaluation \\ Period}} & \multicolumn{10}{c}{Package Managers} \\
    \cline{6-15}
    & & & & & PyPi & NPM & RubyGems & Maven & Cargo & Packagist & NuGet & Go & CPAN & CRAN \\
    \midrule
    
    \multirow{4}{*}{\makecell{Package \\ Vulnerability}} 
    & ~\cite{alfadel2023security} & 1,396 & 2,224 & 2006-20 & \ding{51} &  &  &  &  &  &  & &  &  \\
    & ~\cite{zimmermann2019npmThreats} & 609 & - & 2011-18 &  & \ding{51} &  &  &  &  &  & &  &  \\
    & ~\cite{duan2021supplyChain} & 339 & 1M & n/a & \ding{51} & \ding{51} & \ding{51} &  &  &  &  & &  &  \\
    & ~\cite{zahan2023softwareSecurity} & 5,916 & 958,547 & NA-2022 & \ding{51} & \ding{51} &  &  &  &  &  & &  &  \\
    \midrule
    
    \multirow{4}{*}{\makecell{Dependency \\ Network}} 
    & ~\cite{decan2017dependencyIssues} & 0 & 449,518 & 2011-16 &  & \ding{51} & \ding{51} &  &  &  &  & &  & \ding{51} \\
    & ~\cite{decan2019dependencyEvolution} & 0 & 830,000 & 2012-17 &  & \ding{51} & \ding{51} &  & \ding{51} & \ding{51} & \ding{51} &  & \ding{51} & \ding{51} \\
    & ~\cite{kikas2017dependencyStructure} & 0 & 253,896 & 2005-16 &  & \ding{51} & \ding{51} &  & \ding{51} &  &  & &  &  \\
    \midrule

    \multirow{3}{*}{\makecell{Dependency \\ Vulnerability}} 
    & ~\cite{decan2018npmImpact} & 400 & 610,000 & 2012-17 &  & \ding{51} &  &  &  &  &  & &  &  \\
    & ~\cite{zerouali2022vulnerabilitiesImpact} & 2,874 & 867,290 & 2011-20 &  & \ding{51} & \ding{51} &  &  &  &  & &  &  \\
    & ~\cite{pashchenko2018vulnerableDependencies} & n/a & 200 & n/a &  &  &  & \ding{51} &  &  &  & &  &  \\
    \midrule
    \makecell{Package \\ Vulnerability} & Our Study & \textbf{31,267} & \textbf{14,675} & \textbf{2017-25} & \ding{51} & \ding{51} & \ding{51} & \ding{51} & \ding{51} & \ding{51} & \ding{51} & \ding{51} &  &  \\
    \end{tabular}
    }
    
\end{table*}

%% file: arxiv/02-2-RelatedWork.tex
\section{Background and Related Work}
In this section, we provide a background on the vulnerability disclosure process and describe three main aspects of the security and trustworthiness of OSS projects. 
In particular, we discuss dependency networks, dependency network vulnerabilities, and package-level vulnerabilities. We then discuss our work, focusing on the longitudinal data about reported vulnerabilities and emerging trends in OSS over the past years. 
Table~\ref{tab:related_work} presents the focus of related studies compared to our work. In the following, we describe different categories and briefly discuss the contributions of each study.

\subsection{Vulnerability Disclosure Process}

The vulnerability reporting process is vital for security management, involving key systems to identify and classify issues. Common Vulnerabilities and Exposures \textbf{(CVE)}~\cite{cveDefinition}, managed by MITRE~\cite{mitreWebsite}, is a reference list of publicly known vulnerabilities, each assigned a unique identifier. Complementing CVE, Common Weakness Enumeration \textbf{(CWE)}~\cite{cweWebsite} serves as a community-developed taxonomy for identifying software weaknesses. Each weakness in the CWE list is assigned a unique identifier, simplifying the tracking and addressing of these issues across platforms. Lastly, the National Vulnerability Database \textbf{(NVD)}~\cite{nvdWebsite} enhances CVE data with additional information, improving the context and understanding of vulnerabilities.

\subsection{Dependency Network and Vulnerability Analysis}
Numerous studies explore the structure and evolution of dependency networks in open-source projects. Building on this, other research investigates vulnerabilities within these dependency chains.
We categorize these studies into two groups: \textit{Dependency Network} and \textit{Dependency Vulnerability} analyses. In the following, we will describe each category in more detail. 

\noindent \textbf{Dependency Network Studies}
Decan et al.~\cite{decan2017dependencyIssues}, compare the dependency graphs across NPM, CRAN, and RubyGems, analyzing 449,518 packages from 2011 to 2016. They identify structural issues related to transitive dependencies and compare how these issues evolve over time across different ecosystems.
Decan et al.~\cite{decan2019dependencyEvolution} further explore this and examine the evolution of dependency networks in Cargo, CPAN, CRAN, NPM, NuGet, Packagist, and RubyGems between 2012 and 2017, analyzing approximately 830,000 packages. The authors investigate how the structure and complexity of these networks change over time, providing insights into the growth and maintenance challenges of different ecosystems.
Kikas et al.~\cite{kikas2017dependencyStructure}, study the NPM, RubyGems, and Crates ecosystems from 2005 to 2016, analyzing 253,896 packages. They explore the structural properties and evolution of package dependency networks, highlighting differences and similarities among ecosystems and discussing implications for dependency management tools. 

\noindent \textbf{Dependency Vulnerability Studies}
Decan et al.~\cite{decan2018npmImpact}, present an empirical study of nearly 400 security reports over a six-year period in the NPM ecosystem, which includes over 610,000 JavaScript packages. The authors analyze how and when vulnerabilities are discovered and fixed, and the extent to which they affect other packages within the ecosystem, considering the severity of vulnerabilities and the presence of dependency constraints.
Zerouali et al.~\cite{zerouali2022vulnerabilitiesImpact} conduct an empirical analysis of vulnerabilities reported in the NPM and RubyGems ecosystems between 2011 and 2020, covering 2,874 vulnerabilities in 867,290 packages. They examine how vulnerabilities propagate through dependency networks and assess the impact on both direct and transitive dependents, providing insights into the security risks associated with package dependencies.
The Pashchenko et al.~\cite{pashchenko2018vulnerableDependencies} study focuses on the Maven ecosystem, analyzing the risks associated with vulnerable dependencies. The authors propose methods to assess the actual impact of vulnerabilities by considering factors such as usage popularity and the presence of alternative safe versions, aiming to provide actionable information for software development companies regarding their library dependencies.

Our study does not focus on the dependency networks of vulnerable packages, as prior research has extensively covered this area. Instead, we focus on the vulnerabilities and CWEs across ecosystems, which aligns with the scope of our work. Dependency chain analysis, while valuable, requires an in-depth investigation beyond our current objectives.
 
\subsection{Package Vulnerability Analysis}

Several studies have examined vulnerabilities' nature, frequency, and impact in different ecosystems. This subsection highlights notable research efforts investigating the identification, evaluation, and trends of security vulnerabilities in popular ecosystems.

Alfadel et al.~\cite{alfadel2023security} conducted an empirical study of 1,396 vulnerability reports affecting 698 Python packages in the PyPi ecosystem, covering the period from 2006 to 2020. Their analysis revealed that the discovery and remediation times for vulnerabilities vary significantly, with some vulnerabilities remaining unaddressed for extended periods. The study also highlighted that a substantial number of vulnerabilities are introduced in early versions of packages and persist across releases. 

Zimmermann et al.~\cite{zimmermann2019npmThreats} examined the NPM ecosystem, focusing on its dependency network and associated security risks. Analyzing 609 publicly known security issues from 2011 to 2018, they found that the connected nature of NPM introduces several weak spots. Specifically, installing an NPM package introduces an implicit trust in numerous third-party packages and maintainers, creating a surprisingly large attack surface. The findings suggest that the interconnectedness of the NPM ecosystem can lead to widespread security vulnerabilities.

Zahan et al.~\cite{zahan2023softwareSecurity} investigated the correlation between software security practices and the prevalence of vulnerabilities in open-source packages across the PyPi and NPM ecosystems. Analyzing 5,916 vulnerabilities affecting 958,547 packages up to 2022, they assessed various security measures, such as the use of automated tools, adherence to security guidelines, and community engagement. Their study concluded that packages following robust security practices tend to have fewer reported vulnerabilities, highlighting the effectiveness of proactive security measures in reducing risks.
Zahan et al.~\cite{zahan2022weak} continue the analysis of metadata of 1.63 million NPM packages and proposed signals of security weaknesses.

Our study provides a comprehensive view of vulnerabilities, investigating ten programming languages, a scope unmatched by any prior work, which have typically focused on analyzing a smaller number of ecosystems. Notably, none of the previous studies have included an analysis of C and C++ package vulnerabilities, a gap that our research addresses. Additionally, the vulnerability report datasets analyzed in earlier works are significantly smaller. We analyze 31,267 vulnerability reports from 14,675 packages, making our vulnerability report dataset five times larger than the existing vulnerability analysis in prior studies.
We have listed the prior work in Table~\ref{tab:related_work} to demonstrate what has been the focus of open-source ecosystem study in the past and what differentiates our work from previous studies.

%% file: arxiv/03-Methodology.tex
\section{Research Questions}
\label{sec:rq}

We focus on vulnerabilities reported until 2025, including recent data points for open-source security due to significant shifts and increasing reported CVEs. Most prior studies focus on data from 2018 or earlier. However, we have seen substantial changes in vulnerability patterns in 2022, alongside a significant increase in the number of published packages and reported vulnerabilities. Our analysis addresses these gaps, providing a timely and necessary exploration of new trends and concerns. We address three main research questions that drive our study, designed to uncover key insights into open-source security and provide actionable findings for developers and security practitioners.

\noindent \textbf{RQ1: How have vulnerabilities evolved over time in different ecosystems?} Identifying emerging flaws and vulnerabilities is critical to the development and testing of modern defense solutions such as fuzzing tools, vulnerability scanners, and program analysis techniques. 
This question focuses mainly on trends in open-source software packages and vulnerabilities in published packages across different ecosystems. We investigate how the introduction of new packages across different environments has contributed to the emergence of specific forms of vulnerability. We examine how the introduction of new packages and the overall growth in ecosystems (e.g., NPM, PyPI, Maven) contribute to the emergence of vulnerabilities. We look at Common Vulnerabilities and Exposures (CVEs) compared to the overall growth in the number of packages. 

\textit{Key Finding:} Our analysis reveals that the growth rate of vulnerabilities significantly outpaces package growth across ecosystems, with an average annual increase of 98\% in vulnerable packages compared to 25\% growth in total packages, indicating a deteriorating security posture in open-source ecosystems.

\noindent \textbf{RQ2: How do vulnerabilities persist and concentrate across different ecosystems?} Understanding the distribution of vulnerabilities provides useful insights into ecosystem security dynamics. This research question examines two critical aspects: How long vulnerabilities persist in packages before being addressed (vulnerability lifespan), and how vulnerabilities are concentrated within ecosystems. We analyze vulnerability lifespan trends to understand whether improved community engagement is the reason behind the growth in the number of reported vulnerabilities or not. Additionally, we investigate vulnerability concentration patterns to determine whether security issues are seen in a few critical packages or distributed broadly across ecosystems. This analysis helps identify whether certain packages become repeated targets and informs risk assessment strategies for dependency management.

\textit{Key Finding:} Despite the increasing number of vulnerability reports over time, our study reveals a critical trend: vulnerability lifespans are increasing across all platforms, with an average of 85\% increase for the duration of the study across different ecosystems. This indicates that while more vulnerabilities are being discovered, they persist longer in the software supply chain. Furthermore, we observe significant variations in vulnerability concentration, with ecosystems like Composer showing high concentration (4.92 vulnerabilities per vulnerable package) versus NPM showing broader distribution (1.75 vulnerabilities per package).

\noindent \textbf{RQ3: What are the distinct vulnerability patterns and attack vectors across ecosystems, particularly regarding malicious packages?} Different ecosystems exhibit unique vulnerability characteristics shaped by their design principles and developer practices. This research question explores CWE distribution patterns to identify both common threats and ecosystem-specific vulnerabilities. We analyze the prevalence and trends of different CWEs across major ecosystems from 2017 to 2025, investigating which vulnerabilities are observed in multiple languages and which are specific to individual ones. Additionally, we examine the emerging threat of intentionally malicious packages (CWE-506), which represents a fundamental shift from accidental vulnerabilities to deliberate attacks. We investigate the distribution, attack vectors, and evolution of malicious packages to understand this growing supply chain threat.

\textit{Key Finding:} Our analysis reveals that while certain vulnerabilities like Cross-Site Scripting (CWE-79) and Path Traversal (CWE-22) are observed across different ecosystems, some are deeply tied to language characteristics, such as memory management issues in C/C++ or prototype pollution in JavaScript. Most significantly, we identify an alarming concentration of malicious packages (CWE-506) in NPM and PyPI, where 48.58\% and 13.92\% of all vulnerabilities represent intentional attacks rather than accidental flaws.

%% file: arxiv/04-Data_Collection.tex
\section{Data Collection}
The pipeline integrates two primary data sources of reported vulnerabilities. GitHub Advisory Database \cite{githubadvisory_database}, and Snyk.io \cite{snykWebsite}. These databases are particularly relevant to the research questions due to their structured and detailed categorization of vulnerabilities.
Additionally, we use Libraries.io~\cite{librariesioWebsite} and GitHub.com~\cite{github_com} to collect metadata from the source code and retrieve the information for each repository. Below, we describe each data source, then we demonstrate how we integrated all the gathered information into a comprehensive queryable data source for our research evaluation.

\input{diagrams/tab2}

\vspace{-0.7em}
\subsection{Vulnerability Reports}

\noindent \textbf{GitHub Advisory Database.}
The GitHub Advisory Database~\cite{githubadvisory_database} is a publicly accessible repository that provides security advisories for open-source software projects.
This advisory uses the Open Source Vulnerability format (OSV)~\cite{open_vuln_format} to generate its vulnerability reports. This format includes information such as the package name, ecosystem, severity, affected versions, and remediation steps, as well as references to CVE and CWE identifiers. The Advisory Database offers two distinct datasets for security incidents, covering data from 2017 to the present. \textit{GitHub-reviewed advisories}, which at the time of this study, include over 21,000 reports. And \textit{Unreviewed advisories}, containing more than 242,000 reports. The GitHub-reviewed advisories are security vulnerabilities that have been mapped to packages in the supported languages. GitHub reviews each advisory in this dataset for validity and ensures that they have a full description, and contain both ecosystem and package information. This comprehensive data might not be available for the unreviewed reports. In this study, we only collected the data on the reviewed advisories to ensure that we work on reliable data.
The full advisory dataset is provided by GitHub in a public repository~\cite{githubadvisory_database}. We extracted a total of 17,356 records from the advisory database from eight different ecosystems for our studied programming languages.

\noindent \textbf{Snyk.io}
Snyk Vulnerability Database~\cite{snykWebsite} is another well-supported repository of security vulnerabilities for open-source projects. 
Snyk database aggregates data from its own disclosed vulnerabilities, in addition to other public sources such as the National Vulnerability Database (NVD), community contributions, and its proprietary research team. Snyk specializes in identifying vulnerabilities across a wide range of ecosystems, including all the ones in our study, providing detailed information such as vulnerability descriptions, ecosystems, affected versions, and severity ratings. Prior work has utilized Snyk database for code analysis\cite{kluban2022measuring,sushma2023detect}, Docker vulnerability analysis \cite{kim2023johnny}, and development of vulnerability mitigation tools~\cite{noever2023can}. In total, we collected 13,828 reports from the Snyk database.
Detailed information about each platform's data from both data sources is presented in Table~\ref{tab:data_overview}.

\vspace{-0.7em}

\subsection{Package and Repository Information}

\noindent \textbf{Longitudinal Data from Package Managers.}
A key aspect of our study is the availability of historical data for all package managers. Libraries.io~\cite{librariesioWebsite} is a free platform that aggregates publicly available open-source package information from package manager mirrors and registries on the internet. For all major platforms, including those analyzed in our study, Libraries.io provides dedicated profile pages that display the total number of packages available in each ecosystem's registry.
However, Libraries.io does not offer historical data, it only provides the current number of packages. To address this limitation, we used the Wayback Machine~\cite{internetarchive} to retrieve archived snapshots of Libraries.io's website. From these archives, we collected historical data on the total number of packages for each package manager from 2017 to the present. 
We gathered two data points per year—one in January and one in July—enabling us to track the growth of each ecosystem over time.

\noindent \textbf{Metadata on Published Packages.}
We collected the complete list of available versions for any of the reported packages from Libraries.io’s website. The data includes the version number and publication date of 11,316 unique packages. We will utilize this data to analyze the vulnerability lifespan based on the first vulnerable version and the first patched one.

\subsection{Integrated Dataset Generation}

\noindent \textbf{Unified Vulnerability Reports.}
Although the OSV format has been introduced to streamline advisory databases, adoption remains inconsistent. Even among databases that use OSV, multiple formats exist. For example, affected versions of a CVE are sometimes presented as \texttt{events.introduced}, \texttt{events.fixed}, \texttt{events.last\_affected}, and in other cases with arrays of versions. To address this, we adopted a unified format for affected versions, using brackets (\texttt{[]}) and parentheses (\texttt{()}). 
We established a unified data format to integrate the datasets from GitHub Advisory and Snyk. Each vulnerability report was standardized to include the following fields: {\small\texttt{ADVISORY\_ID}, \texttt{SOURCE}, \texttt{PUBLISH\_DATE}, \texttt{CWES}, \texttt{CVES}, \texttt{ECOSYSTEM}, \texttt{PACKAGE\_NAME}, \texttt{VERSION\_RANGE}, \texttt{REFERENCES}.}

\noindent \textbf{Removing Duplicate Reports}
Detecting duplicates between advisories posed challenges due to inconsistent formatting, particularly for vulnerabilities without CVEs. We found 155 duplicate reports without a CVE with similar \texttt{CWE}, \texttt{Package}, \texttt{Ecosystem}, and \texttt{Affected Versions}. We did a manual verification on them to identify 7 duplicates. Overall, we removed 6,391 duplicate reports, ensuring a cleaner and more reliable dataset.
Additionally, publish dates often differed as reports were published on different platforms at different times. When duplicates were identified, we prioritized Snyk reports over GitHub advisories because they typically contained more detailed information.

\noindent \textbf{Filtering Allowed CWEs.}
We removed the reported vulnerabilities with CWEs that are discouraged or prohibited by MITRE. MITRE has flagged these CWEs as too broad or vague for actionable mitigation, discouraged from use. 
We found 6,770 vulnerability reports associated with these CWEs and removed them from our dataset to get a more accurate insights. 
Examples of such cases are CWE-400 (Uncontrolled Resource Consumption) and CWE-200 (Exposure of Sensitive Information to an Unauthorized Actor). The list of discouraged and prohibited CWEs is provided in Appendix~\ref{app:prohibited_cwes}.

\subsection{Generated Dataset}
The generated dataset includes three distinct and useful features:

\begin{itemize}[itemsep=0pt, topsep=3pt, partopsep=0pt, parsep=0pt, left=5pt]
    \item Information on total packages within each package manager and ecosystem.
    \item Unified vulnerability reports with a standardized format, excluding duplicates and including corrected affected versions.
    \item Package metadata including their vulnerability reports, affected versions, CVEs, and CWEs.
\end{itemize}

We generated a total of 31,267 vulnerability reports, 444 distinct CWEs, and 14,675 unique affected packages from 2017 to 2025. 
A summary of the collected data is presented in Table~\ref{tab:data_overview}.
We believe this dataset will significantly aid future researchers in studying open-source vulnerabilities and package-related information. We plan to keep the dataset updated with new vulnerability reports and package manager updates, providing the community with a unified and well-documented resource to support ongoing research.

%% file: diagrams/tab2.tex
\begin{table}[t]
\centering
\caption{Overview of the collected vulnerability reports from Github Advisory and Snyk database per ecosystem.}
\label{tab:data_overview}
\resizebox{1\linewidth}{!}{%

\begin{tabular}{l|rrrr}
\toprule
Platform & Snyk & \makecell{GitHub \\ Advisory} & \makecell{Unique \\ Reports} & \makecell{Affected \\ Packages} \\
\midrule
C/C++ & 1719 & - & 1,719 & 383 \\
Composer (PHP) & 1,203 & 4,698 & 4,450 & 788 \\
Crates (Rust) & 695 & 1,015 & 1,435 & 575 \\
Go & 1,363 & 2,533 & 1,657 &  \\
Maven (Java) & 1,280 & 6,036 & 5,847 & 2,455 \\
NPM (Node.JS)& 4,740 & 4,248 & 8,237 & 6,245 \\
NuGet (.Net) & 947 & 663 & 1,312 & 581  \\
PyPI (Python) & 1,265 & 3,902 & 3,907 & 1,614 \\
RubyGems (Ruby) & 407 & 944 & 1,191 & 377 \\ \bottomrule
Total & 13,828  & 20,968 & 31,267 & 14,675 \\
\bottomrule

\end{tabular}
}

\end{table}

%% file: arxiv/05-Evaluation.tex
\section{Large-Scale Security Analysis in OSS}
In this section, we aim to answer the research questions raised in Section~\ref{sec:rq}. We analyze data from 10 programming languages: Node.js, PHP (Composer), Python, Java, Ruby, Go, Rust, .NET (C\#), C, and C++. Eight corresponding package managers are also included in this evaluation: NPM, Packagist, PyPI, Maven, RubyGems, Go, Cargo, and NuGet.

\input{arxiv/05-1-Evaluation}
\input{arxiv/05-2-Evaluation}

\input{arxiv/05-3-Evaluation}

\input{arxiv/05-4-Evaluation}

%% file: arxiv/05-1-Evaluation.tex
\subsection{Comparing Vulnerability vs Package Growth}
\label{sec:rq1_growth}

\begin{figure}[t]
    \centering
    \includegraphics[width=\columnwidth]{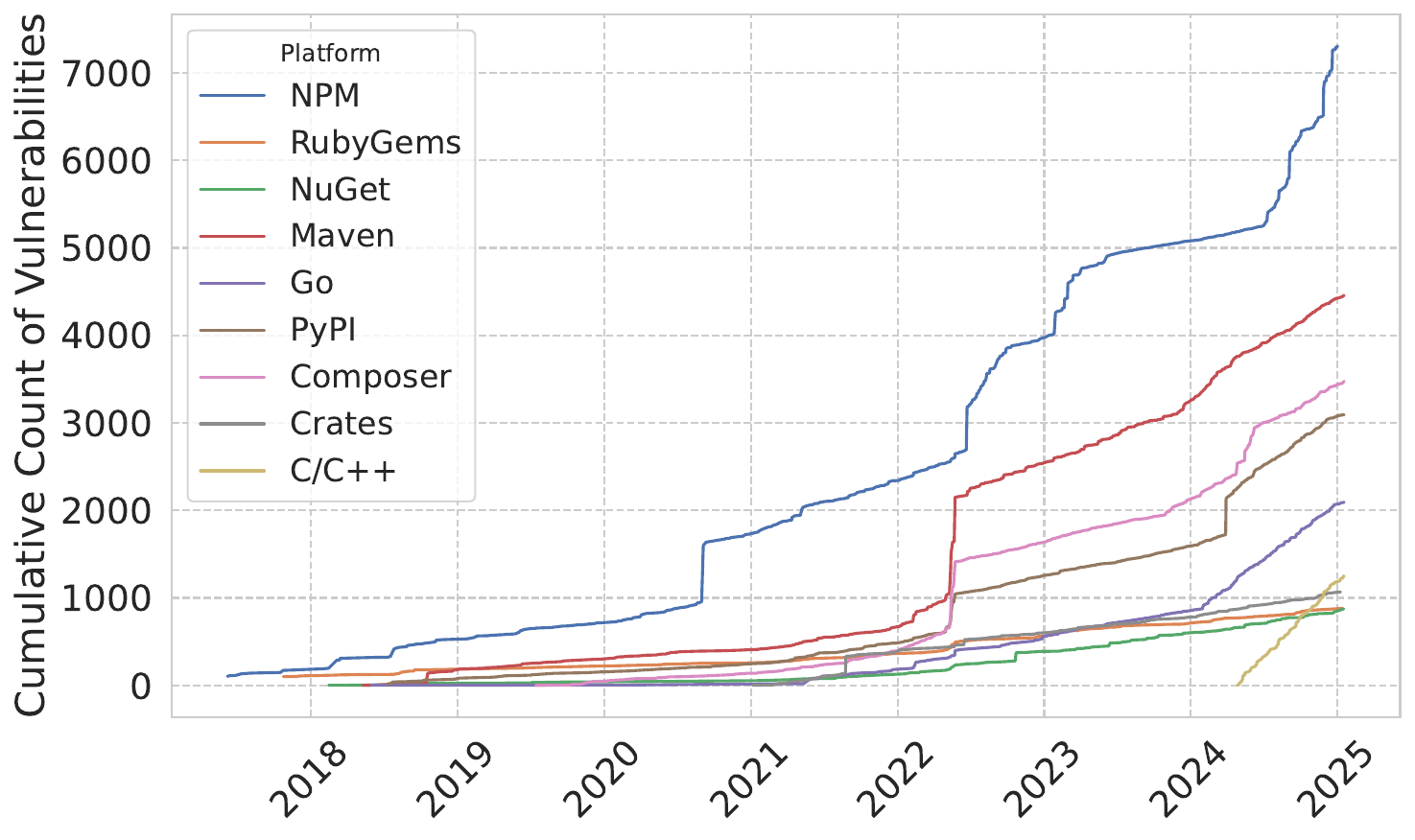}
    \caption{Number of vulnerabilities reported across ecosystems through the years. Reported vulnerabilities increase in every programming language, with an average of over 90 new vulnerabilities in each ecosystem per year.}
    \label{fig:total_vulnerabilities}
\end{figure}

\noindent \textbf{Increase in Total Vulnerabilities Across Languages.}
Alongside the growing number of packages, we observe a significant rise in the total number of reported vulnerabilities across all platforms. As illustrated in Figure~\ref{fig:total_vulnerabilities}, the number of vulnerabilities has significantly increased in every programming language from 2017 to this date. We observe an average of over 90 new vulnerabilities reported in different ecosystems each year. This trend is particularly more significant in fast-growing platforms like NPM, Maven, and Composer, where the surge in both package creation and vulnerabilities has been substantial. In 2022, the number of reported vulnerabilities across different platforms significantly increased, with Maven and Composer reporting 2,497 and 1,592 vulnerabilities, respectively. We do not have sufficient data nor investigated deeply to make a scientific claim about this surge. This increase can be attributed to several factors, including heightened government and regulatory awareness regarding cybersecurity \cite{whitehouse_oss_bill} and high-impact supply chain attacks (such as Log4Shell \cite{Log4Shellibm2024}, MoveIt  \cite{moveit,ciscomoveit}, and Solarwinds \cite{solarwinds} Vulnerabilities)

\begin{figure}[t]
    \centering
    \includegraphics[width=\columnwidth]{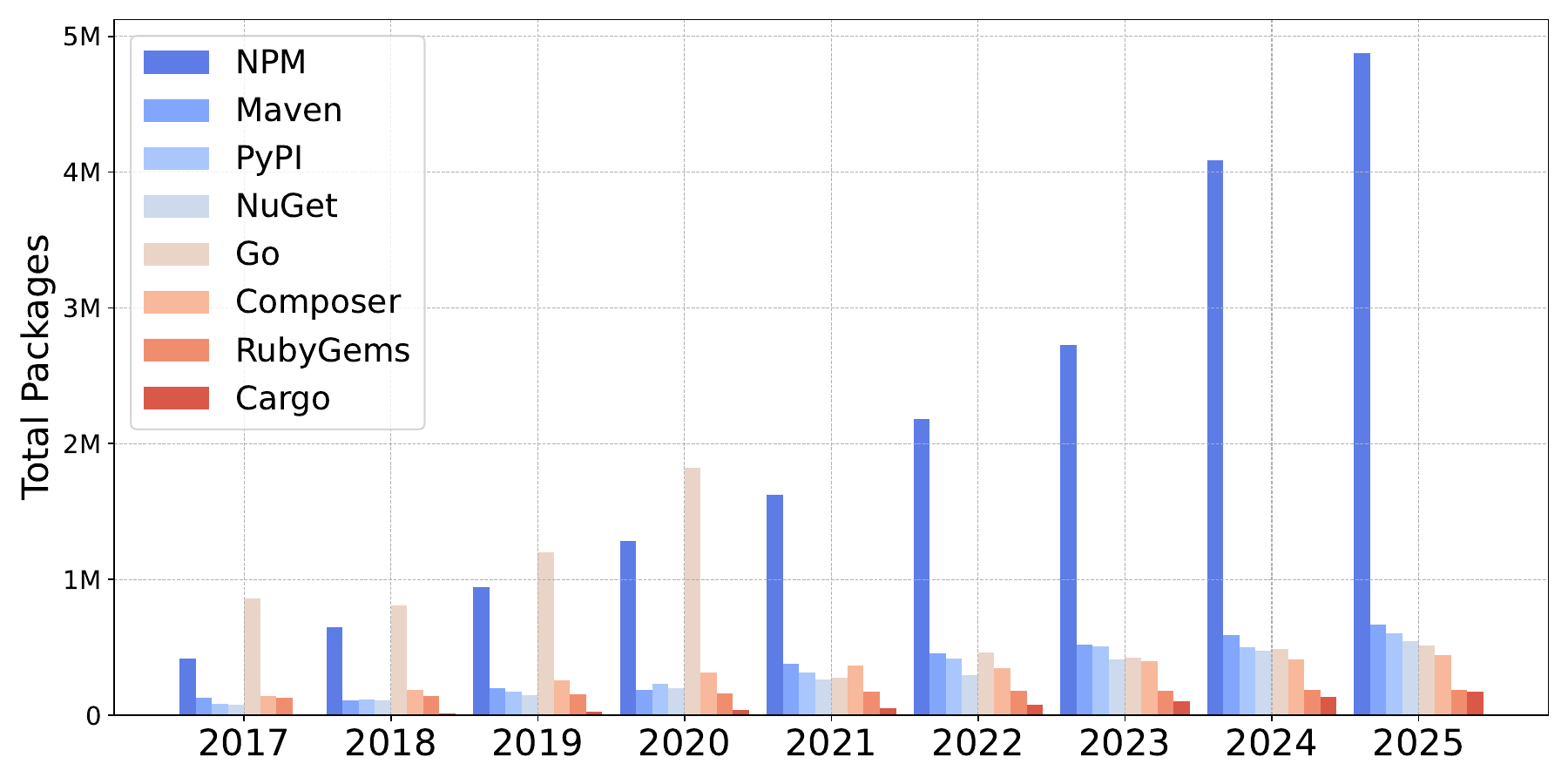}
    \caption{Overview of total packages existing in each package manager. Extracted from libraries.io.}
    \label{fig:ecosystem_overview}
\end{figure}

\noindent \textbf{Increase in Total Number of Packages Across Ecosystems.}
Across all studied platforms, we observe an average of 26.77\% annual increase in the total number of packages.
Figure~\ref{fig:ecosystem_overview} represents the total number of packages in each package manager from 2017 until January 2025. 
We observe the growing popularity and widespread adoption of open-source software, specifically with ecosystems such as NPM, PyPI, and Maven, since they have been experiencing exponential growth over time.
There are exceptions where certain platforms show a temporary decline in package count during specific years, which are probably caused by the removal of outdated and unsupported packages from their registry. For instance, Golang experienced a notable package drop in 2021 following the release of Golang 1.16, which introduced major changes to its module count and dependency management system~\cite{go_mod_ref, go_proxy, maelvls_go111}.
In particular, before Go Modules, Go used a global workspace, defined by the GOPATH environment variable, to manage dependencies where all projects shared the same workspace, making it difficult to manage multiple versions of the same dependency. This led to version conflicts, dependency issues, and challenges in ensuring consistent builds across projects.

\begin{figure}[t]
    \centering
    \includegraphics[width=\columnwidth]{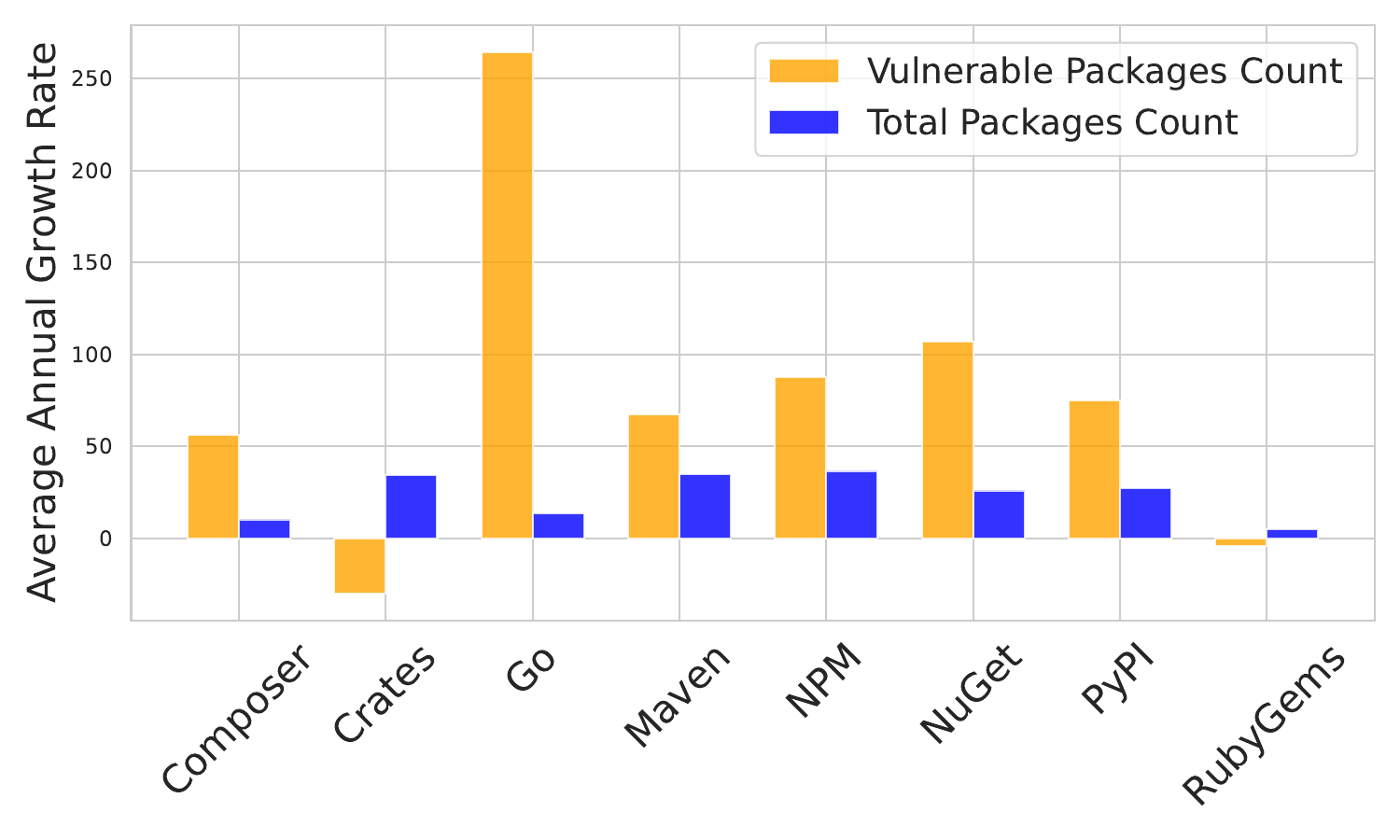}
    \caption{Average annual growth rate of vulnerable packages and total packages by platform. Vulnerabilities have grown at a much faster rate than total packages in nearly all ecosystems.}
    \label{fig:growth_rate_comparison}
\end{figure}

\noindent \textbf{Comparing Growth Rates: Reported Vulnerabilities vs. Total Packages.}
We compare the evolution of vulnerabilities with the growth of package managers over time to examine whether open-source communities have become more secure by analyzing two key factors: the total number of reported vulnerabilities and the total number of packages. We measure and compare the growth rates of these two variables to determine whether security practices have caused the open-source software supply chain to become more secure or whether the popularity and expansion are making them a better target for attackers.
We define the growth rate of vulnerabilities and packages as the year-over-year percentage change in their respective counts.
Figure~\ref{fig:growth_rate_comparison} illustrates the comparative growth rates of vulnerabilities and packages across all measured ecosystems over the years.

Our analysis reveals that the growth rate of vulnerabilities has outpaced the growth rate of total packages across almost all ecosystems. We observe an average of 91\% growth in the number of vulnerable packages compared to 25\% growth in the total number of packages each year. This suggests that while the open-source software community continues to expand, the rate at which vulnerabilities are discovered is growing even faster, raising a major security concern. 
This can potentially be because of the higher community efforts in finding and reporting vulnerabilities. To investigate this further, later in Section \ref{sec:lifespan}, we analyze the decline in security trends from another point of view (i.e., vulnerability lifespan).

%% file: arxiv/05-2-Evaluation.tex
\vspace{0.4em}
\subsection{Vulnerability Lifespan and Concentration}
\label{sec:vuln_conc}
We examine whether the growing popularity and community size of open-source software improve security or introduce new challenges. Additionally, we investigate whether vulnerabilities are concentrated in a few packages or spread across ecosystems. 
These are important questions to answer because one goal of this study is to compare trends across different ecosystems.

\begin{figure}[t]
    \centering
    \includegraphics[width=\linewidth]{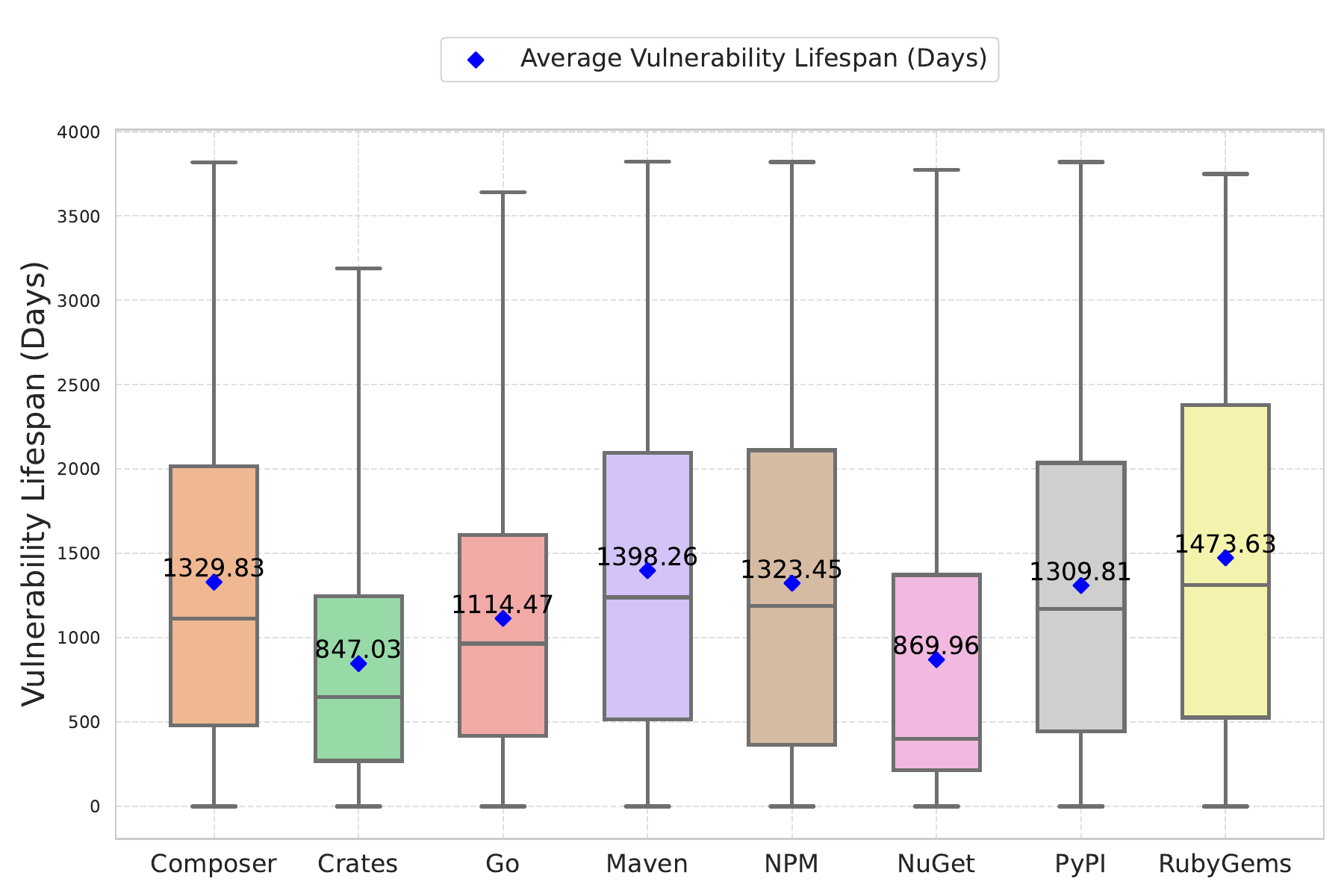}
    \caption{Vulnerability lifespans across ecosystems. Average vulnerability lifespan in different ecosystems varies from 847 to 1,473 days.}
    \label{fig:boxplot_vuln_lifespan}
\end{figure}

\noindent \textbf{Vulnerability Lifespan Across Ecosystems.}
\label{sec:lifespan}
An important aspect of vulnerability analysis is to find out the time that has taken the package maintainers to resolve a reported vulnerability. We define time-to-fix as the duration between the disclosure of a vulnerability and the release date of a patched version. 
However, addressing this question is non-trivial, particularly due to inconsistencies in public data on different repositories as well as the lack of exact disclosure dates for many reported vulnerabilities. 
Moreover, in cases with a disclosure date, we found that the patched version was published before the listed date, making it difficult to provide an accurate analysis of the time-to-fix metric.

In our analysis, we focus on the vulnerability lifespan. We define the lifespan of a vulnerability as the time between the release date of the first vulnerable version of the related package and the release date of the first version in which the vulnerability is resolved.
Figure \ref{fig:boxplot_vuln_lifespan} illustrates the output of 
lifespan analysis among different ecosystems. We do not present any results for C/C++ in this experiment because there are no unified package managers, which makes it impossible to specifically analyze the packages.
We observed that the average vulnerability lifespan across different ecosystems varies from 847 days for Crates to 1,473 days for RubyGems. Comparing these numbers with the version update frequency of each ecosystem, we found out that the frequency of version updates of the vulnerable packages does not correlate with the previous insight. The data suggests that RubyGems and NPM have the highest number of published versions of different packages per day, with an average of 0.06 versions while we saw that it has the largest vulnerability lifespan. The definitions of time-to-fix and lifespan of vulnerabilities are presented in Section \ref{app:terminology} of the Appendix.

\begin{figure}[t]
    \centering
    \includegraphics[width=\linewidth]{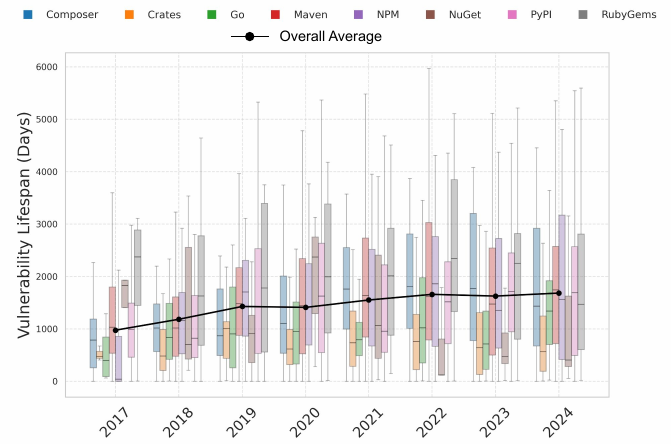}
    \caption{Vulnerability lifespans trend across platforms. While the community efforts have increased, we observe that the average lifespan has increased by 85\% in the past years.}
    \label{fig:boxplot_vuln_lifespan_trend}
\end{figure}

\noindent \textbf{Vulnerability Lifespan Trends.}
In Section \ref{sec:rq1_growth}, we examined the growth in the number of reported vulnerabilities across various ecosystems over the years. One possible explanation for this trend is the increased community efforts involved in vulnerability reporting. To investigate this, we conducted an experiment to assess whether the rise in contributor participation correlates with a reduction in the lifespan of vulnerabilities. 
However, our analysis reveals that greater contributor involvement does not necessarily lead to shorter lifespans. On the contrary, we observe that in most ecosystems, the average vulnerability lifespan has increased in recent years. As shown in Figure \ref{fig:boxplot_vuln_lifespan_trend}, the average lifespan of vulnerabilities across all platforms has grown from 1,056 days to 1,956 days—an increase of approximately 85\%. This finding suggests that despite potentially higher levels of community engagement, vulnerabilities are persisting longer in the software supply chain.

\noindent \textbf{Vulnerability Concentration Across Ecosystems.}
We performed an analysis of the concentration of vulnerabilities in packages across ecosystems. Vulnerability concentration is calculated as the total number of reported vulnerabilities divided by the total number of vulnerable packages.
A higher ratio of vulnerabilities per package indicates a focus on a smaller set of critical or frequently exploited packages, while a ratio closer to one implies a wider spread of vulnerabilities across many packages.

\begin{figure}[t]
    \centering
    \includegraphics[width=\linewidth]{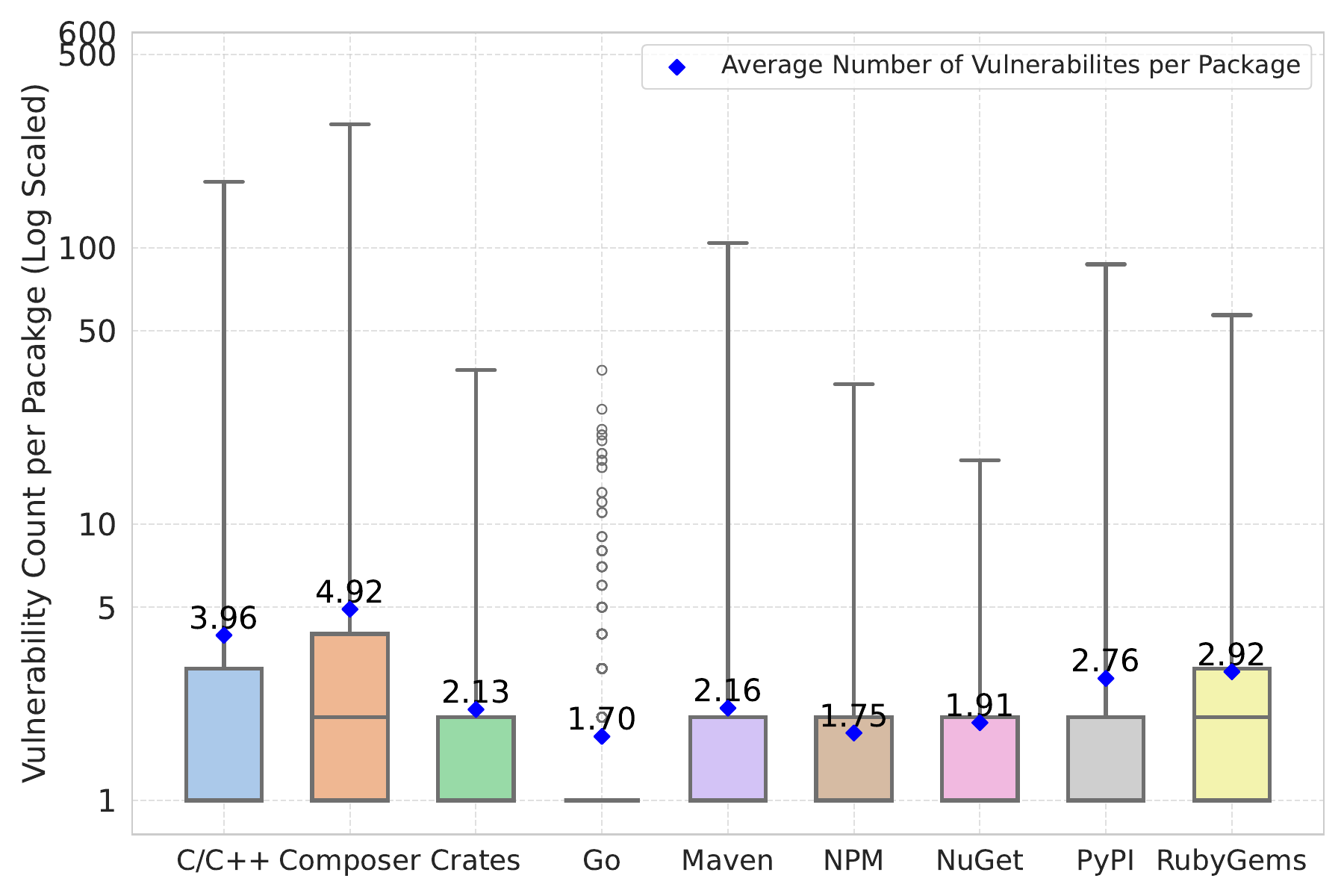}
    \caption{Vulnerability concentration in vulnerable packages across ecosystems.  A higher concentration indicates a focus on a smaller set of critical or frequently exploited packages.}
    \label{fig:boxplot_vuln_per_pkg}
\end{figure}

For example, in Composer, there are an average of 4.92 vulnerabilities per vulnerable package, signaling that certain packages are repeatedly targeted. Similar patterns are observed in C/C++ (3.96) and RubyGems (2.92).
On the other hand, platforms like NPM, with an average of 1.75 vulnerabilities per package, demonstrate a broader distribution of vulnerabilities. Similarly, Go, Maven and PyPI show smaller ratios, suggesting a more even spread of vulnerabilities across packages. These findings, detailed in Figure~\ref{fig:boxplot_vuln_per_pkg}, highlight the diversity of security challenges that ecosystems face. The presented results do not include intentionally malicious packages (Discussed in Section \ref{sec:malicious_packages}) since these packages are reported as a single vulnerability.

We also measured vulnerability concentration among all existing packages in an ecosystem, not only the vulnerable packages.
We did not observe any noticeable changes in the ecosystem rankings based on vulnerability concentration compared to the previous measurement.
Composer is still ranked with the highest overall ratio of 0.86 vulnerabilities per package, while NPM has the lowest at 0.15. Notably, NuGet and NPM both have fewer than 0.25 vulnerabilities per package, indicating significantly low concentration. 

%% file: arxiv/05-3-Evaluation.tex
\subsection{Vulnerability Distribution and Patterns Across Ecosystems.}
\label{sec:rq2_eval}
In this section, we evaluate CWE trends across the studied platforms by analyzing the presence of reported vulnerabilities in different ecosystems. As mentioned in Section \ref{sec:rq}, we aim to identify trends in vulnerability distribution for common CWEs and CWEs specific to ecosystems.

\noindent \textbf{Significance of Top CWEs.}
We performed an analysis on vulnerability patterns across different ecosystems by investigating the most prevalent vulnerability categories. 
\begin{figure}[t]
    \centering
    \includegraphics[width=\linewidth]{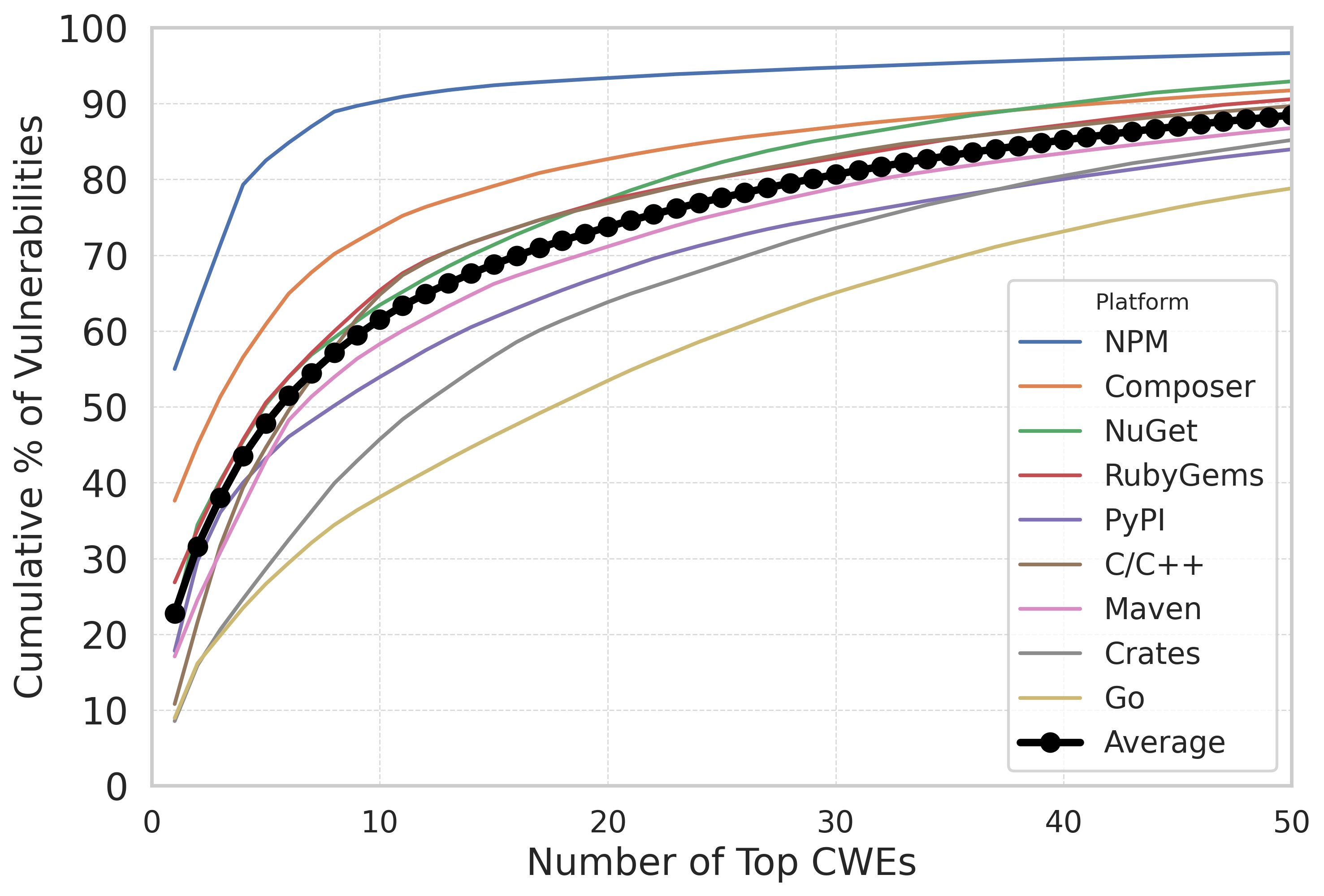}
    \caption{Cumulative graph of vulnerabilities based on top CWEs. A small proportion of CWEs lead to a large number of vulnerabilities. Seven CWEs account for over 50\% of the vulnerabilities across the target platforms.}
    \label{fig:top_cwe_percentages}
\end{figure}
To this end, we generated a cumulative graph that relates the number of vulnerabilities to the number of top CWEs. This visualization helps assess whether vulnerabilities in a given platform are concentrated in a small set of CWE types or spread across a broader range.
Figure \ref{fig:top_cwe_percentages} presents the cumulative percentage of vulnerabilities accounted for by the top 50 CWEs in each ecosystem.
The results reveal a skewed distribution: in ecosystems such as NPM and Composer, a small number of CWE types contribute to a large proportion of vulnerabilities. In contrast, platforms like Go and Crates exhibit flatter curves, indicating a wider distribution of vulnerability types with less concentration in specific CWEs.
These findings highlight the need for ecosystem-specific mitigation strategies. For platforms where vulnerabilities are dominated by a few CWE types, targeted defenses focusing on those CWEs can be particularly effective. On the other hand, ecosystems with a more even distribution of CWEs may require broader, more comprehensive security measures.
In the following, we analyze the most frequently observed CWEs and explore their characteristics to gain deeper insights into the nature of vulnerabilities.

\begin{figure}[t]
    \centering
    \includegraphics[width=\linewidth]{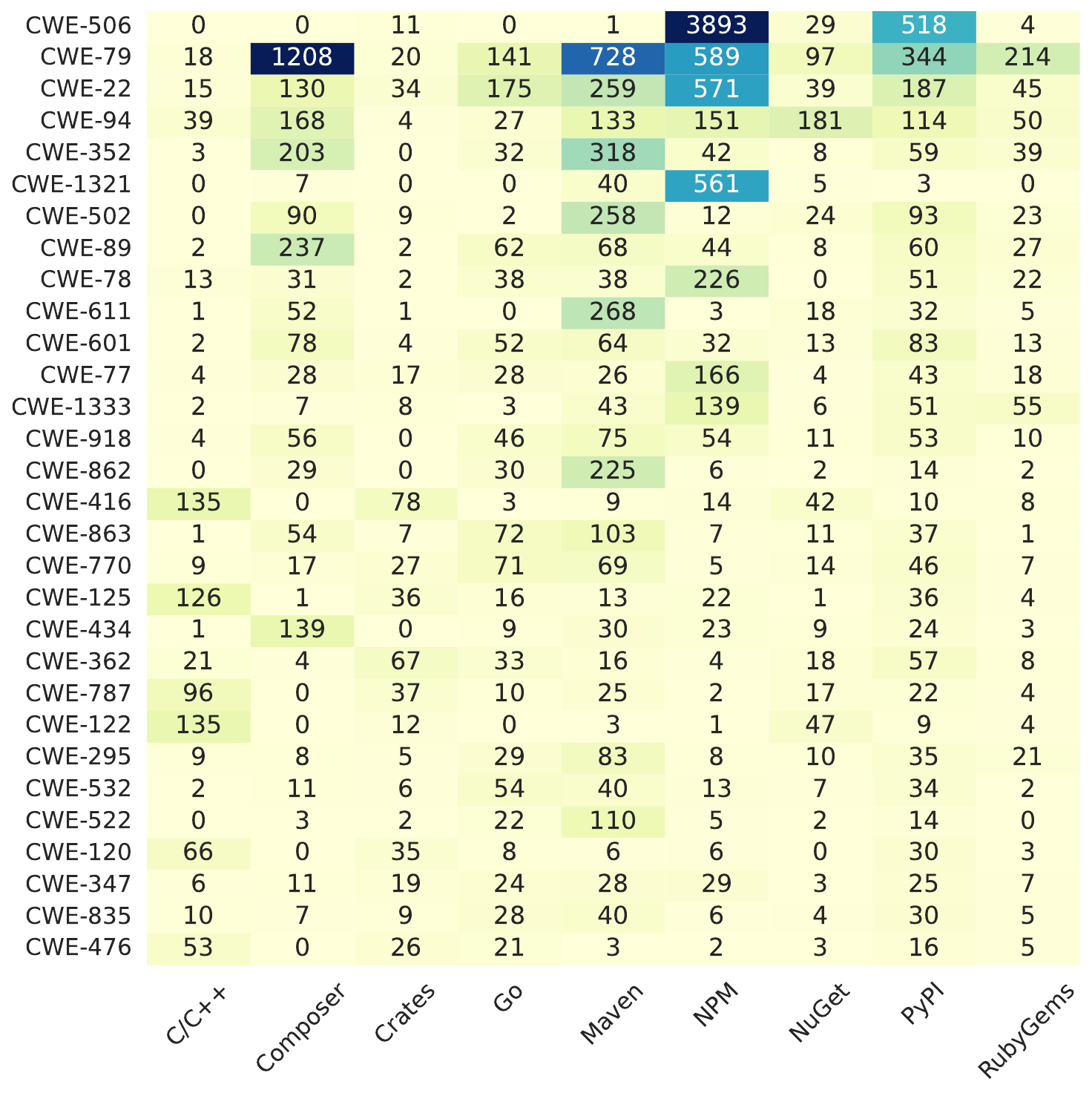}
    \caption{Heatmap of the number of vulnerabilities from top 30 CWEs across different platforms. This suggests that a set of CWEs are common among different programming languages, while others are mainly observed in specific ecosystems.}
    \label{fig:cwe_platform_heatmap}
\end{figure}

\noindent \textbf{Popular CWE Types in Different Ecosystems.} We analyze the distribution of vulnerabilities across programming languages by identifying both common and ecosystem-specific CWEs. 
Figure \ref{fig:top_cwe_percentages} suggests that, on average, only seven CWEs in different ecosystems account for over 50\% of the vulnerabilities. After merging the top CWEs from different platforms, we created a list of the top 28 different CWEs. 
Figure \ref{fig:cwe_platform_heatmap} illustrates a heatmap based on the number of vulnerabilities from a certain CWE in different platforms. This allows us to investigate security trends while highlighting the unique characteristics of certain ecosystems.
In the following, we discuss the insights captured from this analysis.

\textit{General Vulnerability Patterns.} We observe that multiple CWEs, including CWE-79 (Cross-Site Scripting), CWE-22 (Path Traversal), and CWE-94 (Code Injection) are common in several languages over the years and keep growing at a high rate. These common vulnerabilities emphasize fundamental challenges, specifically in web application security and data handling.
  Most web development environments face similar challenges, including user-generated content, improper sanitization, and incorrect output encoding. Table \ref{tab:cwe_summary} in Appendix provides more information for these common CWEs.
  
\textit{Ecosystem-Specific Vulnerability Patterns.}
  Our analysis highlights a set of CWEs that are present in specific ecosystems. For instance, we observe that CWE-1321 (Prototype Pollution) is seen in 560 reports in NPM. This is caused by JavaScript’s prototype inheritance model, where objects can inherit properties from other objects (prototypes). Other examples include CWE-416 (Use After Free) and CWE-362 (Race Condition) which are mainly observed in C/C++ and Crates ecosystems due to their memory handling complexities. 
  Table \ref{tab:cwe_summary} in the Appendix provides examples of specific CWEs in different ecosystems and a brief description for each.

Overall, we observe that specific vulnerability patterns, such as XSS, Path Traversal, and Code Injection are present in multiple ecosystems. However, other classes of vulnerabilities, such as Prototype Pollution or Buffer Overflows, are deeply tied to language characteristics. Each of these vulnerability types potentially requires a different mitigation approach to defend against.
Ecosystem-specific trends emphasize the need for custom analysis rules, threat modeling, and education tailored to the unique risks of each development environment. Figure \ref{fig:cwe_per_ecosystem_trends} in the Appendix provides the trend of top CWEs in each ecosystem.

%% file: arxiv/05-4-Evaluation.tex
\subsection{Analysis of Packages with Malicious Intent (CWE-506)}
\label{sec:malicious_packages}

CWE-506 (Embedded Malicious Code) represents an escalating threat in open-source ecosystems. Unlike other vulnerabilities, which often stem from negligence or poor design, these incidents are rooted in intentional abuse. That is, these vulnerabilities emerge from packages deliberately crafted to be exploited. %
The analysis of the dataset from 2017 to 2025 shows 4,456 reports classified under CWE-506. Notably, these incidents surged from just 38 reports in 2018 to a staggering 2,168 in 2024. %
 This section explores the distribution of these malicious packages, their targeting strategies, and evolving attack vectors. Given their deliberate nature and operational difference, we treat CWE-506 incidents as a separate category, distinct from the other vulnerability trends discussed in Section~\ref{sec:vuln_conc}.

\noindent \textbf{Distribution and Prevalence Across Ecosystems.}
Table~\ref{tab:cwe506_distribution} presents the distribution of CWE-506 vulnerabilities across different ecosystems. It highlights the absolute count of malicious package reports, each ecosystem’s share of the total, and the proportion of its vulnerabilities stemming from intentional attacks.

\input{diagrams/tab3}

Our analysis reveals a high concentration of malicious packages within the NPM and PyPI ecosystems, which together account for 99\% of all CWE-506 reports. The situation is especially concerning in NPM, where nearly half (48.58\%) of all reported vulnerabilities stem from packages designed to inflict harm rather than from unintentional bugs.

Several ecosystem-specific factors contribute to this trend. As noted in prior research~\cite{zimmermann2019npmThreats}, NPM’s low publication barriers, automatic dependency installation, and vast package ecosystem make it particularly attractive for attackers. PyPI also suffers from structural weaknesses that facilitate large-scale abuse. Bagmar et al.~\cite{bagmar2021pypi} describe how PyPI’s packaging model allows malicious packages to propagate easily through dependencies. More recently, Zheng et al.~\cite{zheng2023oscar} found that PyPI lacks robust behavioral vetting and monitoring, leaving the ecosystem vulnerable to complex supply chain attacks. These systemic flaws across both platforms significantly increase their attack surface.

\noindent \textbf{Characteristics of Malicious Packages.}
We investigated all 4,456 malicious packages to understand the tactics attackers use to ensure successful deception and broad reach.

\textit{Package Naming Strategies.}
Our analysis of naming strategies showed that attackers frequently attempt to camouflage their packages by mimicking legitimate libraries or adopting naming conventions that obscure their intent. We found that 3,171 packages (71.2\%) use long names with more than 10 characters, and 2,999 (67.3\%) include dashes in the name, these tactics are likely meant to appear consistent with conventional naming practices. Interestingly, 525 packages (11.8\%) use the scoped format (e.g., \texttt{@org/package}), while 117 packages (2.6\%) adopt very short names (less than 5 characters), perhaps to look core or internal. Additionally, 1,155 packages (25.9\%) use names that closely resemble those of widely adopted libraries, highlighting the significant role of typosquatting and impersonation in attacker strategies. Note that these categories are not mutually exclusive.

\textit{Version Targeting Patterns.}
We also examined versioning patterns and found that attackers typically favor broad compatibility to maximize exposure. Our investigations show that 1,753 packages (39.3\%) target all versions using wildcards such as `\texttt{*}`, aiming for maximum compatibility. Meanwhile, 2,681 packages (60.2\%) target specific versions, indicating a more controlled or environment-specific deployment. Only 22 packages (0.5\%) employ version ranges. This data suggests that attackers typically avoid precise versioning strategies and instead opt for broad compatibility to maximize exposure to more users.

\noindent \textbf{Attack Vectors and Techniques.}
We observed three dominant attack vectors in packages classified under CWE-506: typosquatting, dependency confusion, and installation hook exploitation.

\textit{Typosquatting Attacks.}
Typosquatting attacks exploit human error by publishing packages with names that closely resemble popular libraries~\cite{ohm2020backstabber}. For example, attackers deployed variants such as \texttt{lodahs} (instead of \texttt{lodash}) and \texttt{jquery.js} (instead of \texttt{jquery}). We observed a significant spike in typosquatting attacks in March 2024 flooded PyPI with nearly 200 such malicious packages, all released on the same day. These packages mimicked well-known libraries, including \texttt{pygaqme} (instead of \texttt{pygame}) and \texttt{tensoflouw} (instead of \texttt{tensorflow}).

\textit{Dependency Confusion Attacks.}
Dependency confusion represents a sophisticated supply chain attack where adversaries publish malicious packages to public repositories using names that match private, internal packages used by target organizations~\cite{birsan2021dependency, snyk2022npmdep}. This attack exploits the package resolution mechanism, where the package manager may prioritize public repositories over private ones, or developers may inadvertently install public packages when intending to use internal ones. Our dataset contains 525 scoped packages using the \texttt{@org/package} format, with manual analysis through a sample among these packages, we were able to confirm that there existed various dependency confusion attacks with examples that include \texttt{@ibm-ptc/greet-me} (targeting IBM's private package ecosystem)~\cite{snyk2024ibm}, \texttt{@swiggy-private/analytics} (targeting the food delivery company Swiggy's internal analytics infrastructure)~\cite{snyk2024swiggy}. These attacks are particularly dangerous because they target enterprise environments where the compromise of internal tooling can lead to widespread organizational breaches.

\textit{Installation Hook Exploitation.}
Installation hook exploitation represents a direct and immediate attack vector in the NPM ecosystem, abusing legitimate package lifecycle scripts to execute malicious code during package installation. NPM's package.json specification includes several lifecycle hooks such as \texttt{preinstall} and \texttt{postinstall}, which were designed to allow packages to perform necessary setup tasks like compiling native extensions or configuring environment variables~\cite{snyk2022npmdep}. However, these hooks provide attackers with a powerful mechanism to execute arbitrary code with the same privileges as the user installing the package, often without explicit user consent or awareness. Examples from our dataset include \texttt{postinstall-dummy}~\cite{snyk2022postinstall}, which explicitly advertises its malicious installation script functionality. These attacks are dangerous because they execute immediately upon installation, before developers have an opportunity to review the package contents.

In this section, we covered three more common categories of CWE-506 attacks observed in our dataset. However, 
CWE-506 covers a more diverse scenarios that may not necessarily fall into these categories, but they are still tagged as a CWE-506
Notably, our analysis reveals that even trusted, well-known packages can become vectors for CWE-506 when maintainers deliberately inject malicious code into specific versions. One such case is discussed in the following:

\noindent \textbf{\texttt{node-ipc} Protestware (2022).}  
In March 2022, the maintainer of the popular \texttt{node-ipc} module in NPM, which has more than 600K weekly downloads, introduced geolocation-based file corruption logic in versions 10.1.1 and 10.1.2, targeting users in Russia and Belarus as a political statement against the invasion of Ukraine. The package contained malicious code, that targeted users with IP located in Russia or Belarus, and overwrited their files with a heart emoji~\cite{snyk2022nodeipc}. This incident, later removed in version 10.1.3, marked one of the most prominent examples of "protestware". It illustrates that even trusted packages can become vectors for CWE-506 when maintainers deliberately inject malicious code. This issue was assigned CVE-2022-23812.

Unlike other vulnerabilities, which often stem from negligence or poor design, these incidents are rooted in intentional abuse. The sheer volume, concentration, and strategic sophistication of malicious packages in NPM and PyPI demand dedicated detection mechanisms, supply chain hardening, and ecosystem-wide monitoring.

%% file: diagrams/tab3.tex
\begin{table}[t]
\centering
\caption{Distribution of CWE-506 vulnerabilities across ecosystems. Embedded Malicious Code is mainly reported in NPM and PyPI, with much fewer instances in other ecosystems.}
\label{tab:cwe506_distribution}
\begin{tabular}{lrrr}
\hline
\toprule
Ecosystem & \makecell{CWE-506 \\ Count} & \makecell{\% of All \\ CWE-506} & \makecell{\% of Ecosystem's \\ CWEs} \\
\hline
NPM & 3,893 & 87.37\% & 48.58\% \\
PyPI & 518 & 11.62\% & 13.92\% \\
NuGet & 29 & 0.65\% & 2.33\% \\
Crates & 11 & 0.25\% & 0.86\% \\
RubyGems & 4 & 0.09\% & 0.36\% \\
Maven & 1 & 0.02\% & 0.02\% \\
Others & 0 & 0.00\% & 0.00\% \\
\hline
\textbf{Total} & \textbf{4,456} & \textbf{100.00\%} & \textbf{-} \\
\hline
\end{tabular}
\end{table}

%% file: arxiv/06-Discussion.tex
\section{Discussion}
\noindent \textbf{The Role of Package Managers.}
Our findings reveal that Go's vulnerability growth has been the highest among the studied ecosystems. However, Go has also made significant strides in security, such as with Go Modules, which replaced GOPATH to improve dependency management by removing deprecated and less-maintained packages. In contrast, Rust's Crates.io is the only ecosystem with negative vulnerability growth. This may reflect the security nature of Rust or its lower popularity compared to more commonly targeted ecosystems.
We further observe that the Go community has actively addressed common issues such as  CWE-20 (Improper Input Validation) and CWE-22 (Path Traversal), both of which were among the top five vulnerability types in their ecosystem. However, recent community efforts have led to effective solutions, as seen in the growing family of safe Go libraries~\cite{google_safe_golang}. These libraries, such as \texttt{SafeText}, \texttt{SafeOpen}, and \texttt{SafeArchive} address common security issues in Go, including input validation and path traversal. This suggests that while vulnerabilities in Go have risen, the community has become more proactive, constantly developing defenses and keeping the ecosystem lively.
That said, several strategies can be further improved to enhance the efficacy of package managers. For instance, establishing clearer guidelines for package deprecation, promoting secure default configurations, and sharing more accurate guidelines for package submission can potentially assist developers in maintaining packages more securely. Overall, package managers play a critical role in reducing risks by encouraging secure defaults and actively maintaining packages.

\noindent \textbf{Limited Usage of Existing Protection Protocols.}
Despite ongoing efforts to strengthen the security of open-source software through mechanisms such as automated vulnerability scanning, secure dependency management tools, and transparent builds such as Provenance~\cite{npm_provenance} in the NPM registry, we observe that these protections remain largely underutilized. Although provenance was introduced over a year ago, it has not gained widespread adoption, even among high-risk packages like \texttt{rc}\cite{npmrc}, \texttt{COA}\cite{npmcoa}, and \texttt{ua-parser}~\cite{npmuaparser}. Notably, these same packages were referenced in Provenance’s launch materials, yet they still do not enable the feature.
To better understand the adoption landscape, we examined the NPM's monthly top 100 most-downloaded packages from \texttt{npmleaderboard.org} and found that only 12\% had enabled Provenance. We observe that widely used packages such as \texttt{express} have not adopted the security mechanism. This gap between the availability of protection tools and their actual use leaves many repositories exposed to preventable attacks. The findings emphasize the need for stronger incentives, improved defaults, or tighter integration of these tools into standard development workflows to promote adoption.

\section{Conclusion}
In this paper, we provide a longitudinal analysis of security challenges in 10 different programming languages, revealing an alarming 98\% annual increase in reported vulnerabilities—nearly four times the 25\% growth rate of OSS packages. 
We also reveal the increase in lifespan of vulnerabilities, indicating a potential decline in the security of the ecosystems.
Additionally, our findings highlight that a small number of Common Weakness Enumerations (CWEs) are responsible for the majority of reported vulnerabilities, suggesting that targeted mitigation could significantly improve overall OSS security. 
Furthermore, we discuss the alarming presence of intentionally malicious packages -- particularly in the NPM and PyPI ecosystems -- by analyzing their characteristics and attack vectors.
These insights highlight an urgent need to rethink how we govern and secure open-source ecosystems -- calling for stronger review processes, more effective automated detection tools, greater transparency in package publishing, and a more proactive, community-driven approach to managing software supply chain threats.

%% file: arxiv/99-Appendix.tex
\subsection{Terminology Definitions.}
\label{app:terminology}
Below, we provide the formulas and definitions for each term used in our study to ensure clarity and precision.

\noindent \textbf{Vulnerability Lifespan}
The vulnerability lifespan measures the duration a vulnerability persists in a package before being patched. Since a single package may have multiple affected version ranges, we calculate the total lifespan by summing the durations of all such ranges. This provides a comprehensive view of how long vulnerabilities remain unresolved across different versions.
\[
\scriptsize
\sum_{i=1}^{n} \big( \text{First Patched Version Release Date}_i - \text{First Affected Version Release Date}_i \big)
\]
Here, \(n\) represents the number of affected version ranges for a given package.

\noindent \textbf{Time-to-Fix}
Time-to-Fix measures the duration between the disclosure of a vulnerability and the release of a patched version of the affected package. It is calculated by subtracting the disclosure date of a report from the release date of the first unaffected version. This metric highlights how quickly ecosystems respond to disclosed vulnerabilities.

\subsection{Rise and Fall of Top CWEs}
\begin{figure}[h]
    \centering
    \includegraphics[width=0.9\linewidth]{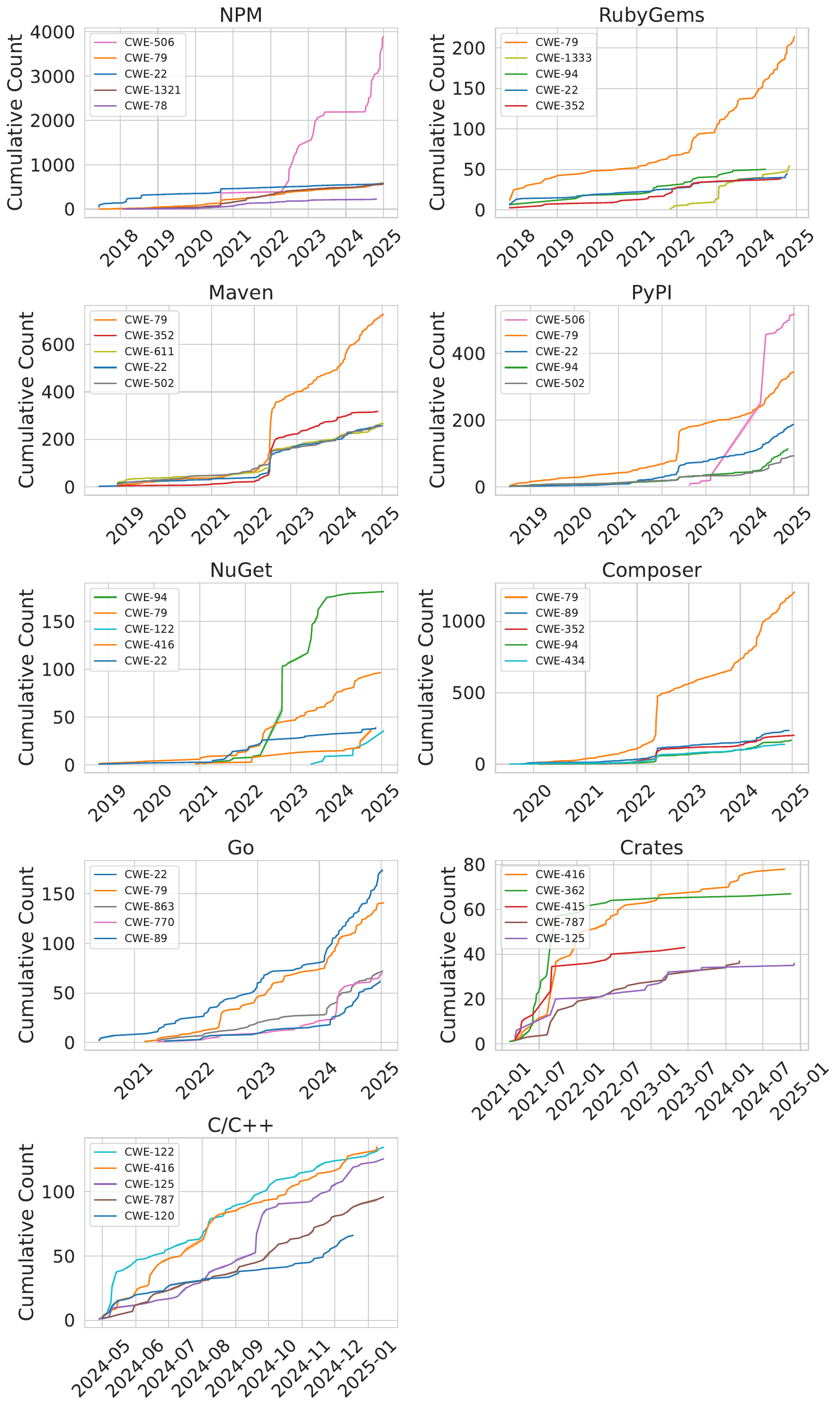}
    \caption{Top five most reported CWEs in each ecosystem over time.}
    \label{fig:cwe_per_ecosystem_trends}
\end{figure}

Figure \ref{fig:cwe_per_ecosystem_trends} illustrates different ecosystem’s top five most reported CWEs over time. These CWEs have either become more prominent or less frequent due to improvements in ecosystem security practices.

\subsection{Discouraged and Prohibited CWEs}
Table \ref{tab:cwe-codes} includes the CWEs that are marked as discouraged or prohibited to use by MITRE.

\label{app:prohibited_cwes}

\begin{table*}[htbp]
\centering
\scriptsize
\caption{Discouraged and Prohibited Common Weakness Enumeration (CWE) and Their Description}
\begin{tabular}{|l|l|l|l|}
\hline
\textbf{CWE ID} & \textbf{Description} & \textbf{CWE ID} & \textbf{Description} \\
\hline
CWE-19 & Data Processing Errors & CWE-200 & Exposure of Sensitive Information to an Unauthorized Actor \\
CWE-20 & Improper Input Validation & CWE-255 & Credentials Management Errors \\
CWE-74 & Improper Neutralization of Special Elements in Output & CWE-264 & Permissions, Privileges, and Access Controls \\
CWE-75 & Failure to Sanitize Special Elements into a Different Plane & CWE-265 & Privilege / Sandbox Issues \\
CWE-118 & Incorrect Access of Indexable Resource ('Range Error') & CWE-269 & Improper Privilege Management \\
CWE-119 & Improper Restriction of Operations & CWE-274 & Improper Handling of Insufficient Privileges \\
CWE-138 & Improper Neutralization of Special Elements & CWE-275 & Permission Issues \\
CWE-284 & Improper Access Control & CWE-285 & Improper Authorization \\
CWE-287 & Improper Authentication & CWE-300 & Channel Accessible by Non-Endpoint \\
CWE-310 & Cryptographic Issues & CWE-311 & Missing Encryption of Sensitive Data \\
CWE-330 & Use of Insufficiently Random Values & CWE-345 & Insufficient Verification of Data Authenticity \\
CWE-372 & Incomplete Internal State Distinction & CWE-391 & Unchecked Error Condition \\
CWE-400 & Uncontrolled Resource Consumption & CWE-435 & Improper Interaction Between Multiple Entities \\
CWE-438 & Behavioral Problems & CWE-610 & Externally Controlled Reference to a Resource in Another Sphere \\
CWE-657 & Violation of Secure Design Principles & CWE-662 & Improper Synchronization \\
CWE-664 & Improper Control of a Resource Through its Lifetime & CWE-665 & Improper Initialization \\
CWE-668 & Exposure of Resource to Wrong Sphere & CWE-680 & Integer Overflow to Buffer Overflow \\
CWE-682 & Incorrect Calculation & CWE-690 & Unchecked Return Value to NULL Pointer Dereference \\
CWE-692 & Incomplete Cleanup & CWE-693 & Protection Mechanism Failure \\
CWE-697 & Incorrect Comparison & CWE-703 & Improper Check or Handling of Exceptional Conditions \\
CWE-707 & Improper Neutralization & CWE-755 & Improper Handling of Exceptional Conditions \\
CWE-788 & Access of Memory Location After End of Buffer & CWE-834 & Excessive Iteration \\
CWE-840 & Business Logic Errors & CWE-1047 & Modules with Circular Dependencies \\
CWE-1056 & Invocation of a Control Element at a Deep Layer & CWE-1068 & Inconsistency Between Implementation and Documented Design \\
CWE-1103 & Use of Platform-Dependent Third Party Components & CWE-1118 & Insufficient Documentation of Error Handling Techniques \\
CWE-1119 & Excessive Use of Unconditional Branching & CWE-1125 & Excessive Attack Surface \\
\hline
\end{tabular}
\label{tab:cwe-codes}
\end{table*}

\subsection{Common and Ecosystem-Specific CWEs}

Table \ref{tab:cwe_summary} includes examples for common and ecosystem-specific CWEs.

\begin{table*}[t]
\centering
\scriptsize
\renewcommand*{\arraystretch}{1.8}
\caption{Common Vulnerability Patterns and Ecosystem-Specific CWEs Across Software Platforms.}
\label{tab:cwe_summary}
\begin{tabular}{l l p{9cm}}
\toprule
\textbf{CWE} & \textbf{Ecosystem(s)} & \textbf{Description} \\
\midrule
\makecell[l]{CWE-79\\(Cross-Site Scripting)~\cite{CWE79}} 
& \makecell[l]{Composer, Maven\\NPM, PyPI} 
& XSS remains a major security concern in web applications, especially dominating Composer (1208), Maven (728), NPM (589), and PyPI (344). \\

\makecell[l]{CWE-22\\(Path Traversal)~\cite{CWE22}} 
& \makecell[l]{NPM, Maven\\PyPI} 
& Path traversal enables attackers to access files outside limited boundaries. NPM (571), Maven (259), and PyPI (187) show the highest number of reports from this category. \\

\makecell[l]{CWE-94\\(Code Injection)~\cite{CWE94}} 
& \makecell[l]{NuGet, Composer\\NPM, Maven, PyPI} 
& Code injection is reported with high frequency in NuGet (181), Composer (168), NPM (151), and Maven (133), followed by PyPI (114). This indicates a high risk in dynamic server-side applications, especially those that handle user input. \\

\midrule

\makecell[l]{CWE-1321\\(Prototype Pollution)~\cite{CWE1321}} 
& NPM 
& Over 560 reports of prototype pollution in NPM highlight the necessity for improved validation mechanisms in libraries that perform object handling. \\

\makecell[l]{CWE-122\\(Heap Buffer Overflow)~\cite{CWE122}} 
& C/C++ 
& The top reported CWE for C/C++ is heap-based buffer overflow (135 reports). This reflects the challenges of manual memory management, leading to memory corruption issues. \\

\makecell[l]{CWE-416\\(Use After Free)~\cite{CWE416}} 
& C/C++, Crates 
& Memory usage after it has been freed leads to undefined behavior or security vulnerabilities. This is mostly observed in C (135) and Crates (78). \\

\makecell[l]{CWE-787\\(Out-of-Bounds Write)~\cite{CWE787}} 
& \makecell[l]{NuGet} 
& This is mostly seen in packages where memory safety practices are crucial. The trends show that this CWE type has decreased in NuGet, likely due to better memory safety practices such as the introduction of Index and Range types in C\# 8 \cite{microsoft_csharp_range_operator_2024}.\\

\makecell[l]{CWE-362\\(Race Condition)~\cite{CWE362}} 
& \makecell[l]{C/C++, Crates} 
& This has been a prominent vulnerability in C and Crates in the past years. However, the reports on this CWE have become less frequent, reflecting improved concurrency handling.\\

\bottomrule
\end{tabular}

\end{table*}